\documentclass[aps,pra,twocolumn,groupedaddress,showpacs,superscriptaddress]{revtex4-2}
\usepackage{amssymb,amsmath,graphicx,epstopdf,hyperref,amsthm,enumitem,color}
\setlist[itemize]{wide=0pt, leftmargin=11pt, labelwidth=6pt, align=left}

\begin{document}
\interfootnotelinepenalty=10000
	
\title{Comparing Singlet Testing Schemes}

\author{George Cowperthwaite}
\affiliation{Centre for Quantum Information and Foundations, DAMTP, Centre for
	Mathematical Sciences,
	University of Cambridge, Wilberforce Road, Cambridge, CB3 0WA, United Kingdom}		
\author{Adrian Kent}
\affiliation{Centre for Quantum Information and Foundations, DAMTP, Centre for
	Mathematical Sciences,
	University of Cambridge, Wilberforce Road, Cambridge, CB3 0WA,
        United Kingdom}
\affiliation{Perimeter Institute for Theoretical Physics, 31 Caroline
  Street North, Waterloo, ON N2L 2Y5, Canada.
\email{A.P.A.Kent@damtp.cam.ac.uk} }

\begin{abstract}
  We compare schemes for testing whether two parties share a two-qubit
  singlet state.  The first, standard, scheme tests Braunstein-Caves (or CHSH)
  inequalities, comparing the correlations of local
  measurements drawn from a fixed finite set against the quantum
  predictions for a singlet.
  The second, alternative, scheme tests the correlations of local
  measurements, drawn randomly from the set of those that are
  $\theta$-separated on the Bloch sphere, against the quantum
  predictions.  We formulate each scheme
  as a hypothesis test and then evaluate the test power in a number of
  adversarial scenarios involving an eavesdropper altering or
  replacing the singlet qubits.  We find the `random measurement' test to be
  superior in most natural scenarios.
\end{abstract}
	
\maketitle

\section{Introduction}\label{sec1} Many quantum information protocols
require two parties (Alice and Bob) to share a two-qubit singlet state
\begin{equation}
|\Psi^-\rangle=\frac{1}{\sqrt{2}}(|01\rangle-|10\rangle),
\end{equation} where Alice holds the first qubit and Bob the second.
Common examples of such two-party protocols include teleportation
\cite{Bennett1993}, summoning tasks \cite{kent2012quantum,kent2013no,hayden2016summoning,adlam2016quantum} and other forms of
distributed quantum computing (e.g. \cite{vaidman2003instantaneous}), 
entanglement-based key distribution protocols (e.g. \cite{Ekert1991}), 
communication and information processing between collaborating agents
in some protocols for position verification and position-based cryptography (e.g. \cite{KMS11,K11.1,buhrman2014position}), and 
in relativistic quantum bit commitment (e.g. \cite{kent2011unconditionally}).

The two parties should be confident they share singlets, both to ensure
the protocol will execute as intended and to
preclude the possibility of an adversarial third-party having
interfered with the system for their own advantage.  A natural method
for distinguishing singlets from other quantum states is to measure a
quantity for which the singlet attains a unique maximum.
Common examples are the expressions in the CHSH \cite{Clauser1969} and
Braunstein-Caves \cite{Braunstein1990} inequalities.
These have the additional advantage that they test Bell nonlocality.
They can thus detect any adversarial attack that replaces the singlet qubits with classical physical systems programmed to
produce deterministic or probabilistic results in response
to measurements, since these can be modelled by local hidden variables. 
In this paper, we examine a quantity that has the same properties but has not
previously been studied as a singlet test, the (anti-)correlation of outcomes of
random measurements separated by a fixed angle $\theta \in ( 0 , \pi/3 )$,
and compare it to schemes
derived from the CHSH and Braunstein-Caves inequalities (e.g. \cite{McKague2012,Valcarce2019,Bancal2015,Yang2014}).

The singlet testing schemes we examine in this paper only require 
both parties to accurately perform projective measurements.
Unlike singlet purification
schemes \cite{bennett1996purification}, they do not need
quantum computers or quantum memory. 
This makes them potentially advantageous when users' technology
is limited, or more generally when single-qubit measurements
are cheap compared to multi-qubit operations.

We assume that Alice and Bob are separated, and the singlet is created
either by Alice or by a source separated from both, before each qubit is transmitted to the respective party.
An adversarial third party may intercept the qubits
during transmission and alter them, either to obtain information or to disrupt the protocol.
We refer to her as `Eve', but emphasize that we are interested in protocols beyond
key distribution and that her potential interference need not necessarily involve eavesdropping. 

We compare the power of our proposed singlet
testing schemes against four commonly studied attacks.
These do not represent the full range of possible adversarial action,
but illustrate why random measurement testing schemes can be advantageous
in a variety of scenarios: 
\begin{enumerate} \item Single
qubit intercept-resend attack: Eve intercepts Bob's qubit, performs a
local projective measurement, notes the outcome and sends the
post-measurement state on to Bob. This could occur in a setting where
Alice creates the singlet and transmits a qubit to Bob.  \item
Bipartite state transformation: Eve intercepts both qubits and
performs some quantum operation on them, replacing the singlet by some
other two-qubit state that is sent to Alice and Bob.  \item LHV
replacement: Eve replaces the singlet with some non-quantum system
chosen so that Alice's and Bob's measurement outcomes are determined
by local hidden variables instead of quantum entanglement.  \item
Noisy quantum channel: described by a physically natural noise model
(hence a special case of scenario 2, if we consider the noise as due
to Eve). This alters the singlet state as it is transmitted to Alice
and Bob.  \end{enumerate}

The advantages of these various attacks for Eve, in disrupting or obtaining
information from the protocol, will depend on the context.
We assume each offers Eve some potential advantage and 
focus on the extent to which Alice and Bob can detect the 
attacks.

We consider two different types of scheme which Alice and Bob may use
to test the purported singlet:
\begin{itemize}
\item Braunstein-Caves test: testing the Braunstein-Caves inequality
  \cite{Braunstein1990} with a specific set of $N$ measurement
  choices for which the singlet uniquely induces the maximum
  violation \cite{Wehner2006}. We often focus in particular on the $N=2$ case, 
  the CHSH inequality \cite{Clauser1969}, for which 
  self-testing schemes have been extensively studied
  (e.g. \cite{McKague2012,Valcarce2019,Bancal2015,Yang2014}).
	\item Random measurement test: Alice and Bob choose random
          local projective measurements constrained to have a fixed
          separation angle on the Bloch sphere \cite{Kent2013} and
          calculate the anti-correlation of their measurement outcomes.
          For a wide range of angles, this is uniquely maximized by the singlet.

\end{itemize}

The intuition, which we test and quantify,
is that the random measurement test may generally be more efficient than Braunstein-Caves, as it tests anti-correlations for the same set of axis separations $( \pi/2N)$,
but chooses axes randomly over the Bloch sphere, giving Eve less information about the test measurements and hence less scope to tailor her attack to minimize its detectability.
It can also be applied for any $\theta$, not just the discrete set of the form $\{\pi/2N:N\in\mathbb{N},N>1\}$.

We first describe these schemes and analyse their efficiency.
We discuss their feasibility in the final section.   

\section{Braunstein-Caves singlet test}\label{sec2}
\subsection{Preliminaries}
Alice and Bob wish to test whether they share a two-qubit singlet. It
is possible that they instead share a not necessarily quantum system governed by a
local hidden variable theory (see section \ref{sec4c}) or a more complex
quantum system with further degrees of freedom.  However we start by
assuming the parties are confident they share a (potentially mixed)
two-qubit state $\rho_{AB}$. To start the test, Alice and Bob use an
authenticated channel to fix a parameter $N\geq2$ and uniformly
randomly generate a projective measurement pair $(A,B)$ from the set
\begin{equation}\label{mmset}
	\{(a_k,b_k)\}_{k=0}^{N-1}\cup\{(a_{k+1},b_k)\}_{k=0}^{N-2}\cup\{(a_0,b_{N-1})\},
\end{equation}
where
\begin{gather}
	a_k=\left\{|m_{k\pi/2N}\rangle,|m_{k\pi/2N+\pi/2}\rangle\right\}, \label{am}\\
	b_k=\left\{|m_{(2k+1)\pi/4N}\rangle,|m_{(2k+1)\pi/4N+\pi/2}\rangle\right\}, \label{bm}
\end{gather}
with $|m_\theta\rangle=\cos\theta|0\rangle+\sin\theta|1\rangle$ and outcomes are labelled $\{1,-1\}$ respectively.

Next, Alice and Bob perform measurements $A$ and $B$ respectively on their qubit and store their outcomes as $O_{A}$ and $O_{B}$ respectively. They compute the quantity
\begin{equation}
	\hat{C}=\begin{cases}
		-O_{A}O_{B} & \text{ when } ( A,B ) = (a_0,b_{N-1}) \\
		O_{A}O_{B} & \text{ otherwise}
	\end{cases}
\end{equation}
through an authenticated classical channel. We call $\hat{C}$ the `Braunstein-Caves sample', as it possesses properties derived from the Braunstein-Caves inequality \cite{Braunstein1990}.

$\hat{C}$ takes outcomes $\{1,-1\}$, so it follows a shifted Bernoulli distribution. The expected value of $\hat{C}$ resulting from a uniformly random choice of measurement bases $(A,B)$ and application of those measurements is
\small\begin{equation}\label{bcexp}
	E[\hat{C}]=\frac{1}{2N}\left(\sum_{k=0}^{N-1} E[a_k,b_k]+\sum_{k=0}^{N-2} E[a_{k+1},b_k]-E[a_0,b_{N-1}]\right),
\end{equation}
\normalsize
with $E[x,y]$ defined as the expected correlation between Alice and Bob's measurement outcomes for choices $(A,B)=(x,y)$. The expectation is bounded \cite{Wehner2006} for quantum states by
\begin{equation}
	|E[\hat{C}]|\leq\cos\left(\frac{\pi}{2N}\right)
\end{equation}
and the singlet saturates this bound  \cite{Wehner2006}.

As we review below, for the measurement choices defined in (\ref{am}) and
(\ref{bm}), the singlet \textit{uniquely} achieves the minimum expectation
$(-\cos(\pi/2N))$. Any other state thus produces a detectable deviation
in the sample mean of $\hat{C}$, given a large enough sample size,
assuming perfect measurements.

\subsection{Calculating $E[\hat{C}]$ for $\rho_{AB}$}
If Alice and Bob each utilise local projective measurements, a general
combined measurement basis for a Braunstein-Caves test can be described by
\begin{equation}
	\begin{split}
	\{&|m_\theta\rangle|m_\phi\rangle,|m_{\theta+\frac{\pi}{2}}\rangle|m_{\phi+\frac{\pi}{2}}\rangle,\\
	&|m_{\theta}\rangle|m_{\phi+\frac{\pi}{2}}\rangle,|m_{\theta+\frac{\pi}{2}}\rangle|m_{\phi}\rangle\}
	\end{split}
\end{equation}
where $|m_\theta\rangle=\cos\theta|0\rangle+\sin\theta|1\rangle$, with
the first two results corresponding to correlated outcomes and the
final two results corresponding to anticorrelated outcomes.

Define $a_{ijpq}=\langle ij|\rho_{AB}|pq\rangle$ for $i,j,p,q\in\{0,1\}$.
The expected correlation between outcomes of measurement $\{|m_{\theta}\rangle,|m_{\theta+\frac{\pi}{2}}\rangle\}$ on Alice's qubit and $\{|m_{\phi}\rangle,|m_{\phi+\frac{\pi}{2}}\rangle\}$ on Bob's qubit is given by
\begin{equation}\label{bccalc}
	\begin{split}
	&E\left[\{|m_{\theta}\rangle,|m_{\theta+\frac{\pi}{2}}\rangle\},\{|m_{\phi}\rangle,|m_{\phi+\frac{\pi}{2}}\rangle\}\right]\\
	=&P(\text{outcomes same})-P(\text{outcomes differ})\\
	=&2P(\text{outcomes same})-1\\
	=&2\langle m_\phi|\langle m_\theta|\rho_{AB}|m_\theta\rangle|m_\phi\rangle|^2\\
	&+2\langle m_{\phi+\frac{\pi}{2}}|\langle m_{\theta+\frac{\pi}{2}}|\rho_{AB}|m_{\theta+\frac{\pi}{2}}\rangle|m_{\phi+\frac{\pi}{2}}\rangle|^2-1\\
	=&\cos2\theta\cos2\phi(a_{0000}+a_{1111}-a_{0101}-a_{1010})\\
	&+2\cos2\theta\sin2\phi\cdot\text{Re}(a_{0001}-a_{1011})\\
	&+2\sin2\theta\cos2\phi\cdot\text{Re}(a_{0010}-a_{0111})\\
	&+2\sin2\theta\sin2\phi\cdot\text{Re}(a_{0011}+a_{0110}).
	\end{split}
\end{equation}
Thus by utilising values of $\theta,\phi$ corresponding to the measurements in (\ref{am}) and (\ref{bm}), quantity (\ref{bcexp}) evaluates as
\small\begin{equation}\label{BCeval}
	\begin{split}
	E[\hat{C}]=&\frac{1}{2N}\left(\sum_{k=0}^{N-1} E[a_k,b_k]+\sum_{k=0}^{N-2} E[a_{k+1},b_k]-E[a_0,b_{N-1}]\right)\\
	=&\frac{1}{2}\cos\left(\frac{\pi}{2N}\right)(a_{0000}+a_{1111}-a_{0101}-a_{1010})\\
	&+\cos\left(\frac{\pi}{2N}\right)\text{Re}(a_{0011}+a_{0110})\\
	=&\cos\left(\frac{\pi}{2N}\right)\Big(\langle\Phi^+|\rho_{AB}|\Phi^+\rangle-\langle\Psi^-|\rho_{AB}|\Psi^-\rangle\Big),
	\end{split}
\end{equation}
\normalsize
where $|\Phi^+\rangle=\frac{1}{\sqrt{2}}(|00\rangle+|11\rangle)$ and $|\Psi^-\rangle=\frac{1}{\sqrt{2}}(|01\rangle-|10\rangle)$.

Clearly, the minimum of $E[\hat{C}]$ is uniquely attained by the
singlet. 

\section{Random measurement singlet test}\label{sec3}
\subsection{Preliminaries}
Alice and Bob wish to test whether they share a two-qubit
singlet. Again, we start by assuming the parties are confident they share a
(potentially mixed) two-qubit state $\rho_{AB}$.  Alice uniformly
randomly generates a projective qubit measurement
\begin{equation}
\{|\psi_A\rangle,|\psi_A^\perp\rangle\},
\end{equation} 
corresponding to outcomes $\{1,-1\}$ and, likewise, Bob uniformly randomly generates a projective qubit measurement 
\begin{equation}
\{|\psi_B\rangle,|\psi_B^\perp\rangle\},
\end{equation}
from a set with the defining restriction that $|\psi_A\rangle$ and
$|\psi_B\rangle$ must be separated by angle $\theta\in[0,\pi/2]$ on the Bloch
sphere, so that $|\langle\psi_A|\psi_B\rangle|=\cos\theta$. This separation can be achieved in many ways:
for example, Alice and Bob could share a list of pre-agreed
measurements or Bob could delay his measurement choice until Alice has
made and communicated hers. The optimal method of achieving this
depends on the parent protocol within which the shared singlet is
required.   For example, pre-sharing of measurement choices may be reasonable when verifying singlets for use in a teleportation protocol, but may not be
motivated in a key generation scheme, as a one-time pad
could be pre-shared instead with similar resources.

Next, Alice and Bob perform their chosen measurements on their qubit and compute the product of their outcomes through an authenticated classical channel to obtain 
\begin{equation}
	\hat{O}=O_AO_B,
\end{equation}
where $O_A$ and $O_B$ are the outcomes of Alice and Bob's measurements respectively. We call $\hat{O}$ the `random measurement sample'.

$\hat{O}$ takes outcomes $\{1,-1\}$, so it follows a shifted Bernoulli distribution. The expected value of $\hat{O}$ resulting from a uniformly random choice of $\theta$ separated measurement bases on the Bloch sphere and application of those measurements is denoted by $E[\hat{O}]$.

It will be shown that, for this test, the singlet uniquely attains the minimum expectation $-\cos\theta$. Any other state would produce a significant deviation in the sample mean of $\hat{O}$, given a large enough sample size.

\subsection{Relation between $|\psi_A\rangle$ and $|\psi_B\rangle$}
If $|\psi_A\rangle$ and $|\psi_B\rangle$ are separated by angle $\theta$ on the Bloch sphere, they are related by
\begin{equation}
	\begin{split}
	|\psi_B\rangle=&U_{|\psi_A\rangle}P(\alpha)R(\theta/2)|0\rangle,\\
	|\psi_B^\perp\rangle=&U_{|\psi_A\rangle}P(\alpha)R(\theta/2)|1\rangle,
	\end{split}
\end{equation}
for some $\alpha\in[0,2\pi)$, where $U_{|\psi_A\rangle}=|\psi_A\rangle\langle0|+|\psi_A^{\perp}\rangle\langle1|$ is a unitary transformation and
\begin{equation}
P(\alpha)=\begin{pmatrix}
	1&0\\
	0&e^{i\alpha}
\end{pmatrix},\;\;\;\;\;\\R(\theta/2)=\begin{pmatrix}
	\cos(\theta/2)&-\sin(\theta/2)\\
	\sin(\theta/2)&\cos(\theta/2)
\end{pmatrix},
\end{equation}
in the computational basis. Thus
\begin{equation}\label{psib}
	\begin{split}
	|\psi_B\rangle=&\langle0|P(\alpha)R(\theta/2)|0\rangle|\psi_A\rangle+\langle1|P(\alpha)R(\theta/2)|0\rangle|\psi_A^{\perp}\rangle,\\
	|\psi_B^\perp\rangle=&\langle0|P(\alpha)R(\theta/2)|1\rangle|\psi_A\rangle+\langle1|P(\alpha)R(\theta/2)|1\rangle|\psi_A^{\perp}\rangle.
	\end{split}
\end{equation}

\subsection{$E_{AB}[\hat{O}]$ for fixed pair of measurements}
For ease of notation, define the following product states
\begin{equation}
	\begin{split}
	|\psi_{AA}\rangle&=|\psi_A\rangle|\psi_A\rangle\;\;\;\; |\psi_{AA}^{\perp\perp}\rangle=|\psi_A^\perp\rangle|\psi_A^\perp\rangle,\\ 
	|\psi_{A^\perp A}\rangle&=|\psi_A^\perp\rangle|\psi_A\rangle\;\;\;\; |\psi_{AA^\perp}\rangle=|\psi_A\rangle|\psi_A^\perp\rangle,\\
	|\psi_{AB}\rangle&=|\psi_A\rangle|\psi_B\rangle\;\;\;\; |\psi_{AB}^{\perp\perp}\rangle=|\psi_A^\perp\rangle|\psi_B^\perp\rangle.
	\end{split}
\end{equation}

Let $E_{AB}[\hat{O}]$ be the expected measurement correlation for a fixed choice of $|\psi_A\rangle$ and $|\psi_B\rangle$. Then
\begin{equation}
	\begin{split}
	E_{AB}[\hat{O}]&=P[\hat{O}=1]-P[\hat{O}=-1]\\
	&=2P[\hat{O}=1]-1\\
	&=2\langle\psi_{AB}|\rho_{AB}|\psi_{AB}\rangle+2\langle\psi_{AB}^{\perp\perp}|\rho_{AB}|\psi_{AB}^{\perp\perp}\rangle-1.
	\end{split}
\end{equation}
Using the expressions for $|\psi_B\rangle$ in (\ref{psib}) to evaluate each term individually
\begin{equation}
	\begin{split}
		E_{AB}[\hat{O}]=&2\cos^2(\theta/2)\langle\psi_{AA}|\rho_{AB}|\psi_{AA}\rangle\\
		&+2\sin^2(\theta/2)\langle\psi_{AA^\perp}|\rho_{AB}|\psi_{AA^\perp}\rangle\\&
		+2\text{Re}[e^{i\alpha}\sin(\theta)\langle\psi_{AA}|\rho_{AB}|\psi_{AA^\perp}\rangle]\\
		&+2\cos^2(\theta/2)\langle\psi_{AA}^{\perp\perp}|\rho_{AB}|\psi_{AA}^{\perp\perp}\rangle\\
		&+2\sin^2(\theta/2)\langle\psi_{A^\perp A}|\rho_{AB}|\psi_{A^\perp A}\rangle\\&
		-2\text{Re}[e^{i\alpha}\sin(\theta)\langle\psi_{AA}^{\perp\perp}|\rho_{AB}|\psi_{A^\perp A}\rangle]-1.
	\end{split}
\end{equation}

\subsection{$E_{A}[\hat{O}]$ for a fixed Alice measurement}
For a fixed measurement choice for Alice, the expected correlation
$E_{A}[\hat{O}]$ over all Bob's possible measurement choices is found
by integrating over $\alpha$ in $[0,\pi]$. The integrals of
$e^{i\alpha}$ and $e^{-i\alpha}$
vanish over this interval, hence
\begin{equation}\label{expA}
	\begin{split}
	&E_{A}[\hat{O}]=\frac{1}{\pi}\int_0^\pi E_{AB}[\hat{O}]\text{ d}\alpha\\
	=&2\cos^2(\theta/2)\big[\langle\psi_{AA}|\rho_{AB}|\psi_{AA}\rangle+\langle\psi_{AA}^{\perp\perp}|\rho_{AB}|\psi_{AA}^{\perp\perp}\rangle\big]-1\\
	&+2\sin^2(\theta/2)\big[\langle\psi_{AA^\perp}|\rho_{AB}|\psi_{AA^\perp}\rangle+\langle\psi_{A^\perp A}|\rho_{AB}|\psi_{A^\perp A}\rangle\big]\\
	=&\cos\theta\big[2\langle\psi_{AA}|\rho_{AB}|\psi_{AA}\rangle+2\langle\psi_{AA}^{\perp\perp}|\rho_{AB}|\psi_{AA}^{\perp\perp}\rangle-1\big].
	\end{split}
\end{equation}

Note that $|\psi_A\rangle$ can be written as
\begin{equation}
	\begin{split}
	|\psi_A\rangle=&\cos(\omega/2)|0\rangle+e^{i\beta}\sin(\omega/2)|1\rangle,\\
	|\psi_A^\perp\rangle=&\sin(\omega/2)|0\rangle-e^{i\beta}\cos(\omega/2)|1\rangle,
	\end{split}
\end{equation}
for some $\omega\in[0,\pi]$ and $\beta\in[0,2\pi)$, so that 
\begin{equation}
	\begin{split}
	|\psi_{AA}\rangle=&\frac{1}{2}\left(1+\cos\omega\right)|00\rangle+\frac{1}{2}e^{i\beta}\sin\omega|01\rangle\\
	&+\frac{1}{2}e^{i\beta}\sin\omega|10\rangle+\frac{1}{2}e^{2i\beta}\left(1-\cos\omega\right)|11\rangle,\\
	|\psi_{AA}^{\perp\perp}\rangle=&\frac{1}{2}\left(1-\cos\omega\right)|00\rangle-\frac{1}{2}e^{i\beta}\sin\omega|01\rangle\\
	&-\frac{1}{2}e^{i\beta}\sin\omega|10\rangle+\frac{1}{2}e^{2i\beta}\left(1+\cos\omega\right)|11\rangle.
	\end{split}
\end{equation}

For ease of notation, define the quantities 
\begin{equation}
	a_{ijpq}=\langle ij|\rho_{AB}|pq\rangle,
\end{equation}
for $i,j,p,q\in\{0,1\}$. Using these quantities, equation (\ref{expA}) gives
\begin{equation}\label{finalexpA}
	\begin{split}
	E_{A}[\hat{O}]=\cos\theta&\big[2\langle\psi_{AA}|\rho_{AB}|\psi_{AA}\rangle+2\langle\psi_{AA}^{\perp\perp}|\rho_{AB}|\psi_{AA}^{\perp\perp}\rangle-1\big]\\
	=\cos\theta&\bigg[\frac{1}{2}(1+\cos\omega)^2a_{0000}+\frac{1}{2}(1-\cos\omega)^2a_{1111}\\
	&+\frac{1}{2}(1-\cos\omega)^2a_{0000}+\frac{1}{2}(1+\cos\omega)^2a_{1111}\\
	&+\sin^2\omega\cdot a_{0101}+\sin^2\omega\cdot a_{1010}\\
	&+2\sin^2\omega\cdot\text{Re}[a_{0110}]-1\\
	&+f(e^{i\beta},e^{2i\beta},e^{-i\beta},e^{-2i\beta})\bigg]\\
	=\cos\theta&\Big[\cos^2\omega-(\cos2\omega+1)(a_{0101}+a_{1010})\\
	&+2\sin^2\omega\cdot\text{Re}[a_{0110}]\\
	&+f(e^{i\beta},e^{2i\beta},e^{-i\beta},e^{-2i\beta})\Big],
	\end{split}
\end{equation}
where $f$ is a function representing a linear combination of its arguments.

\subsection{Calculating $E[\hat{O}]$}
The expected correlation over all of Alice's possible measurement choices is found by integrating (\ref{finalexpA}) over $\beta$ in $[0,2\pi]$ and over $\omega$ in $[0,\pi]$, with the Jacobian $\sin\omega$ appropriate for integration over a sphere surface. The integral of $e^{in\beta}$ vanishes over the interval $[0,2\pi]$, hence
\begin{equation}\label{RMeval}
	\begin{split}
	E[\hat{O}]=\frac{1}{4\pi}\int_0^{2\pi}\int_0^\pi &E_{A}[\hat{O}]\sin\omega\text{ d}\omega\text{d}\beta\\
	=\frac{1}{2}\cos\theta\int_0^\pi\Big[&\cos^2\omega-(\cos2\omega+1)(a_{0101}+a_{1010})\\
	&+2\sin^2\omega\cdot\text{Re}[a_{0110}]\Big]\sin\omega\text{ d}\omega\\
	=\cos\theta\bigg(\frac{1}{3}-\frac{2}{3}&(a_{0101}+a_{1010})+\frac{4}{3}\text{Re}[a_{0110}]\bigg)\\
	=\cos\theta\bigg(\frac{1}{3}-\frac{4}{3}&\langle \Psi^-|\rho_{AB}|\Psi^-\rangle\bigg),
	\end{split}
\end{equation}
where $\langle \Psi^-|\rho_{AB}|\Psi^-\rangle$ is the fidelity between $\rho_{AB}$ and the singlet state.\\\\ 
Clearly, the minimum of $E[\hat{O}]$ is uniquely attained by the
singlet. 

\section{Comparison of test efficiency}\label{sec4}
We will now describe and compare hypothesis tests for the singlet using (i) Braunstein-Caves samples with parameter N or (ii) random measurement samples with $\theta=\pi/2N$. We link our choice of $\theta$ to $N$ in this way, as this ensures both tests induce equal expected correlations when measuring singlets, allowing for a clear comparison of the effect of deviations. We recall that the test samples both follow shifted Bernoulli distributions
\begin{equation}
	\begin{split}
	\text{BC sample }\hat{C}&\sim2\cdot\text{Bernoulli}(p)-1,\\
	\text{RM sample }\hat{O}&\sim2\cdot\text{Bernoulli}(q)-1,
	\end{split}
\end{equation}
with parameters defined through the expectation values in (\ref{BCeval}) and (\ref{RMeval}) as
\begin{equation}\label{pqdefn}
	\begin{split}
	p&=\frac{1}{2}+\frac{1}{2}\cos\left(\frac{\pi}{2N}\right)\Big(\langle \Phi^+|\rho_{AB}|\Phi^+\rangle-\langle \Psi^-|\rho_{AB}|\Psi^-\rangle\Big),\\
	q&=\frac{1}{2}+\frac{1}{2}\cos\left(\frac{\pi}{2N}\right)\left(\frac{1}{3}-\frac{4}{3}\langle \Psi^-|\rho_{AB}|\Psi^-\rangle\right).
	\end{split}
\end{equation}

If we denote the sample mean of $n$ Braunstein-Caves samples as $\overline{C}$ and the sample mean of $n$ random measurement samples as $\overline{O}$, then they both follow shifted binomial distributions
\begin{equation}\label{samplemeans}
	\begin{split}
		\text{BC sample mean }\overline{C}&\sim\frac{2}{n}\text{B}(n,p)-1,\\
		\text{RM sample mean }\overline{O}&\sim\frac{2}{n}\text{B}(n,q)-1.
	\end{split}
\end{equation}

\subsection{Description of hypothesis tests}\label{sec4a}
We aim to test the hypotheses
\begin{equation}
	H_0:\rho_{AB}=|\Psi^-\rangle\langle\Psi^-| \;\;\text{ v }\;\; H_1:\rho_{AB}\not=|\Psi^-\rangle\langle\Psi^-|.
\end{equation}

If we wish to conduct the test using Braunstein-Caves samples, we generate $n$ samples of $\hat{C}$ (as in section \ref{sec2}) and let the test statistic be $\overline{C}$.

If we wish to conduct the test using random measurement samples, we generate $n$ samples of $\hat{O}$ (as in section \ref{sec3}) and let the test statistic be $\overline{O}$.

Let $\alpha$ be the desired size of the test, defined as the probability the null hypothesis is erroneously rejected when Alice and Bob do in fact share a singlet. The critical region $R$ is a set of values for the test statistic for which the null hypothesis is rejected. For both tests, we wish to define $R$ as
\begin{equation}\label{rdefn}
	R=\left\{x:x>\frac{2z_{\alpha,n}}{n}-1\right\},
\end{equation}
where $z_{\alpha,n}$ is defined as the upper $\alpha$-quantile of a $B\big(n,\frac{1}{2}-\frac{1}{2}\cos\left(\frac{\pi}{2N}\right)\big)$ distribution. However, binomial quantiles can only take discrete values, so we are often unable to select one exactly correponding to $\alpha$. To rectify this, we instead set $z_{\alpha,n}$ to be the smallest integer such that $P\left(X>z_{\alpha,n}|X\sim B\left(n,\frac{1}{2}-\frac{1}{2}\cos\left(\frac{\pi}{2N}\right)\right)\right)<\alpha$ and extend $R$ to a critical decision region $R^+$, where if our test statistic exactly equals $2z_{\alpha,n}/n-1$, we decide to reject the null hypothesis with probability $q$, where $q$ is chosen so that $P\left(X\in R^+|X\sim B\left(n,\frac{1}{2}-\frac{1}{2}\cos\left(\frac{\pi}{2N}\right)\right)\right)=\alpha$.

Through (\ref{samplemeans}), it follows that $P(\overline{C}\in R^+|H_0)=P(\overline{O}\in R^+|H_0)=\alpha$, as $\overline{C}$ and $\overline{O}$ are identically distributed under the null hypothesis. 

The power functions $\pi_{BC}$ and $\pi_{RM}$ for each test describe the probability the null hypothesis is rejected given the density matrix of the state being tested, and are defined using (\ref{pqdefn}),
(\ref{samplemeans}) and (\ref{rdefn}) as
\begin{equation}
	\begin{split}
		\pi_{BC}(\rho_{AB})&=P(X\in R^+|X\sim B(n,p)),\\
		\pi_{RM}(\rho_{AB})&=P(X\in R^+|X\sim B(n,q)).
	\end{split}
\end{equation}

It is clear that \begin{equation} \begin{split}
\pi_{BC}(\rho_{AB})>\pi_{RM}(\rho_{AB})&\iff p>q\\ &\iff
E[\hat{C}]>E[\hat{O}], \end{split} \end{equation} so whether (i) or (ii)
is better at detecting non-singlet states in a given
scenario can be determined by comparing the values of $E[\hat{C}]$ and
$E[\hat{O}]$ associated with testing typical states arising from that
scenario.

For large $n$, the asymptotic power functions are described by the central limit theorem. If $\Phi$ is the cumulative distribution function of a $N(0,1)$ distribution, then as $n\to\infty$
\begin{equation}\label{asymp}
	\begin{split}
		\pi_{BC}(\rho_{AB})&\sim 1-\Phi\left(\frac{\tilde{z}_{\alpha,n}+0.5-np}{\sqrt{np(1-p)}}\right),\\
		\pi_{RM}(\rho_{AB})&\sim 1-\Phi\left(\frac{\tilde{z}_{\alpha,n}+0.5-nq}{\sqrt{nq(1-q)}}\right),
	\end{split}
\end{equation}
where the `+0.5' terms are the appropriate correction for a continuous limit of a discrete distribution and $\tilde{z}_{\alpha,n}$ is the upper $\alpha$-quantile of a $N(np_0,np_0(1-p_0))$ distribution, with $p_0=\frac{1}{2}-\frac{1}{2}\cos\left(\frac{\pi}{2N}\right)$.

\subsection{Comparison for simple intercept-resend attack}\label{sec4b}
Consider the scenario in which Eve manages to intercept Bob's qubit and performs a measurement in basis
\begin{equation}
	\{|\alpha\rangle=\cos\psi|0\rangle+e^{i\beta}\sin\psi|1\rangle,|\alpha^\perp\rangle\},
\end{equation}
before sending the post-measurement qubit on to Bob. As the singlet can be expressed as
\begin{equation}
	|\Psi^-\rangle=\frac{1}{\sqrt{2}}|\alpha\rangle|\alpha^\perp\rangle-\frac{1}{\sqrt{2}}|\alpha^\perp\rangle|\alpha\rangle,
\end{equation}
it is clear that the post-measurement state will be		
\begin{equation}
	\rho_{AB}=\begin{cases}
		|\alpha\rangle\langle\alpha|\otimes|\alpha^\perp\rangle\langle\alpha^\perp| & \text{with probability } 1/2\\
		|\alpha^\perp\rangle\langle\alpha^\perp|\otimes|\alpha\rangle\langle\alpha| & \text{with probability } 1/2.
	\end{cases}
\end{equation}
The expected test samples are independent of Eve's measurement outcome and are calculated in Table \ref{table:1} using (\ref{BCeval}) and (\ref{RMeval}).
\begin{table}[h!]
	\begin{tabular}{ |c|c|c| } 
		\hline
		  & $\rho_{AB}$ & $|\Psi^-\rangle$ \\
		 \hline
		$E[\hat{C}]$ & $\left(-\frac{1}{2}+\frac{1}{4}\sin^2(2\psi)\sin^2(2\beta)\right)\cos\left(\frac{\pi}{2N}\right)$ & $-\cos\left(\frac{\pi}{2N}\right)$ \\
		\hline
		$E[\hat{O}]$ & $-\frac{1}{3}\cos\left(\frac{\pi}{2N}\right)$ & $-\cos\left(\frac{\pi}{2N}\right)$ \\ 
		\hline
	\end{tabular}
\caption{Comparison of expectation values for post measurement state and singlet, with the BC test using parameter $N$ and the RM test using $\theta=\pi/2N$.}
\label{table:1}
\end{table}

These results show that a single qubit intercept-resend attack reduces $E[\hat{O}]$ to $1/3$ of its singlet value, and reduces $E[\hat{C}]$ to between $1/4$ and $1/2$ of its singlet value, depending on the measurement made by Eve.

Eve will choose $(\psi,\beta)$ to achieve a desired balance of minimal
disruption and maximal information gain, hence her choice will depend
on the parent protocol within which Alice and Bob intended to use the
singlet. For example, if BB84 is the parent protocol, it is known
\cite{Dan2009} that the Breidbart basis $(\beta=0,\psi=\pi/8)$ is
optimal for Eve, hence $E[\hat{O}]>E[\hat{C}]$ and the random
measurement test has greater power. More generally, if Eve's priority
is to choose a basis which minimises disruption, the random
measurement test will be more powerful.

\begin{figure}
\includegraphics[scale=0.44]{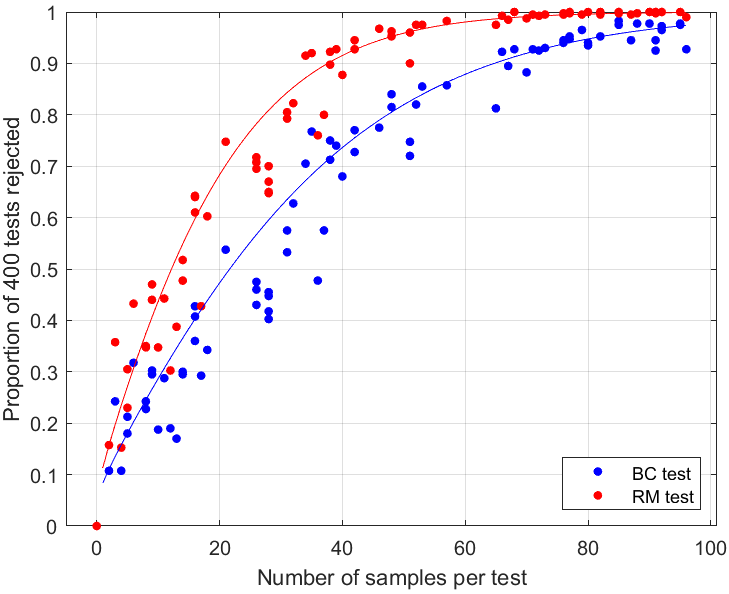}
\centering
\caption{Results when Eve measures in the computational basis. Test parameters $N=2$ and $\theta=\pi/4$. Curves are theoretical asymptotic power functions (\ref{asymp}). Dots are empirical data, representing the proportion of 400 simulated tests leading to a rejection of the null hypothesis at the $1\%$ level.}
\end{figure}

It is clear that choosing $N$ to be as large as possible will maximise the difference between the expected correlations under the null and alternative hypotheses, leading to the more powerful test for both schemes. This implies choosing $\theta=0$ is optimal for the random measurement test in this scenario, hence it is optimal for Alice and Bob to use the same randomly chosen measurements if they know they are testing between a singlet and a post-measurement state. However, it is known \cite{Bell1964} (see section \ref{sec4d}) that for $\theta=0$, the test does not distinguish the singlet from a class of simple LHV models.

\subsection{Comparison for bipartite state transformation attack}\label{sec4c}
Consider the scenario in which Eve intercepts both qubits and manipulates them, so that the singlet is transformed into some other state $\rho_{AB}$, with singlet fidelity
\begin{equation}
	\langle\Psi^-|\rho_{AB}|\Psi^-\rangle=1-\epsilon.
\end{equation}
While we permit any $\epsilon\in[0,1]$, we are particularly interested in small values. The expected test samples are calculated in Table \ref{table:2} using (\ref{BCeval}) and (\ref{RMeval}).
\begin{table}[h!]
	\begin{tabular}{ |c|c|c| } 
		\hline
		& $\rho_{AB}$ & $|\Psi^-\rangle\langle\Psi^-|$ \\
		\hline
		$E[\hat{C}]$ & $\left(-1+\epsilon+\langle\Phi^+|\rho_{AB}|\Phi^+\rangle\right)\cos\left(\frac{\pi}{2N}\right)$ & $-\cos\left(\frac{\pi}{2N}\right)$ \\
		\hline
		$E[\hat{O}]$ & $\left(-1+\frac{4}{3}\epsilon\right)\cos\left(\frac{\pi}{2N}\right)$ & $-\cos\left(\frac{\pi}{2N}\right)$ \\ 
		\hline
	\end{tabular}
\caption{Comparison of expectation values for post transformation state and singlet, with the BC test using parameter $N$ and the RM test using $\theta=\pi/2N$.}
\label{table:2}
\end{table}

The results show that $E[\hat{O}]$ increases with $\epsilon$ at linear
rate $4\cos(\pi/2N)/3$, while $E[\hat{C}]$ increases with
$\left(\epsilon+\langle\Phi^+|\rho_{AB}|\Phi^+\rangle\right)$ at
linear rate $\cos(\pi/2N)$, thus the test of greater power can be
identified by comparing the values of
$\langle\Phi^+|\rho_{AB}|\Phi^+\rangle$ and $\epsilon$. Note that the
orthogonality of Bell states imposes the constraint
$0\leq\langle\Phi^+|\rho_{AB}|\Phi^+\rangle\leq\epsilon$.

When $0\leq\langle\Phi^+|\rho_{AB}|\Phi^+\rangle<\epsilon/3$, we have $E[\hat{O}]>E[\hat{C}]$, and when $\epsilon/3<\langle\Phi^+|\rho_{AB}|\Phi^+\rangle\leq\epsilon$, we have $E[\hat{C}]>E[\hat{O}]$, while for $\langle\Phi^+|\rho_{AB}|\Phi^+\rangle=\epsilon/3$, both tests are equally strong.

\begin{figure}[b]
\includegraphics[scale=0.44]{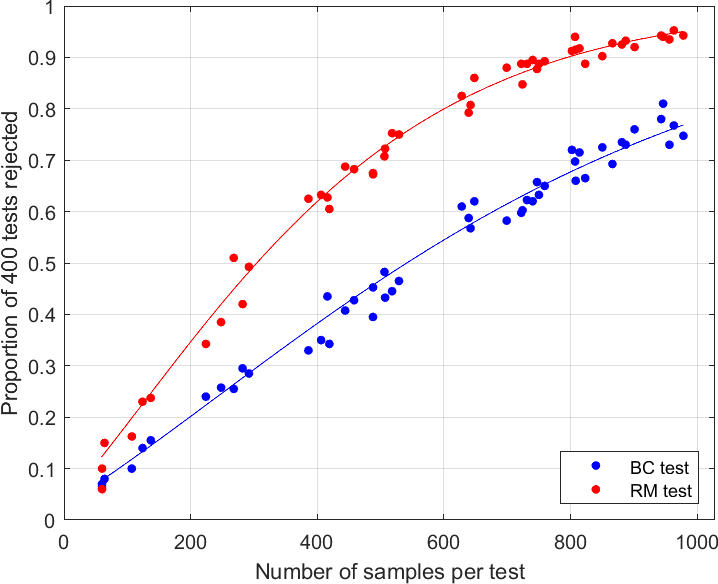}
\centering
\caption{Results when Eve transforms singlets to states with $\langle\Psi^-|\rho_{AB}|\Psi^-\rangle=0.9$ and $\langle\Phi^+|\rho_{AB}|\Phi^+\rangle=0$. Test parameters $N=2$ and $\theta=\pi/4$. Curves are theoretical asymptotic power functions (\ref{asymp}). Dots are empirical data, representing the proportion of 400 simulated tests leading to a rejection of the null hypothesis at the $1\%$ level.}
\end{figure}

As an example, in a scenario where Eve prioritises being as undetectable as possible for a given $\epsilon$, she would choose a transformation with $\langle\Phi^+|\rho_{AB}|\Phi^+\rangle=0$, so the random measurement test would be superior in this case.

One way to overcome the uncertainty in the value of $\langle\Phi^+|\rho_{AB}|\Phi^+\rangle$ is to require Alice and Bob to apply the same randomly chosen unitary operation $U$ to both of their qubits before measurement, without remembering the identity of $U$. This effectively transforms their shared system to a mixed state of form
\begin{equation}
	(1-\epsilon)|\Psi^-\rangle\langle\Psi^-|+\frac{\epsilon}{3}\left(\mathbb{I}-|\Psi^-\rangle\langle\Psi^-|\right)
\end{equation}
as the singlet component remains invarient under a $U\otimes U$ transformation, while the complement becomes maximally mixed. This ensures that both tests are equivalently strong when testing the resulting state, as $\langle\Phi^+|\rho_{AB}|\Phi^+\rangle=\epsilon/3$.

The largest possible choice of parameter $N$ leads to the test of greater power for each type of scheme, much as it did in section \ref{sec4b}.

\subsection{Comparison for LHV replacement attack}\label{sec4d}
Consider the scenario in which Eve intercepts both qubits and replaces them with some not necessarily quantum system where the correlation between Alice and Bob is governed entirely by a local hidden variable (LHV) theory. Using the Braunstein-Caves inequality \cite{Braunstein1990}, the expected value of the Braunstein-Caves sample is bounded for integers $N\geq2$ as
\begin{equation}\label{BCineq}
	|E[\hat{C}]|\leq1-\frac{1}{N}<\cos\left(\frac{\pi}{2N}\right),
\end{equation}
for measurements of a two-sided LHV system. It is also known (Theorem 1 in \cite{Kent2013}) that the expected value of the random measurement sample for $\theta=\pi/2N$ is bounded for integers $N\geq2$ as
\begin{equation}\label{RMineq}
	|E[\hat{O}]|\leq1-\frac{1}{N}<\cos\left(\frac{\pi}{2N}\right),
\end{equation}
for measurements of a two-sided LHV system. The strictly positive difference between correlations resulting from LHV models and singlets implies that both tests can detect when the correlation between Alice and Bob's measurement outcomes is caused by a LHV theory, with a power uniformly bounded for all possible LHV theories.

For $N\geq2$, the optimal parameters for both types of scheme are found by selecting the value of $N$ that maximises the difference between expected singlet correlations and the bound on LHV correlations. This difference is defined in (\ref{BCineq}) and (\ref{RMineq}) as
\begin{equation}\label{gap}
	D(N)=\cos\left(\frac{\pi}{2N}\right)+\frac{1}{N}-1.
\end{equation}
 
As $D(2)\approx0.207$, $D(3)\approx0.199$ and $D'(N)<0$ for $N\geq3$, it follows that $D(N)$ is maximised by $N=2$ over integer inputs greater than 1, providing an optimal minimum bound on test power for both schemes.

This result does not identify which value of $\theta$ leads to the
random measurement test of greatest power for detecting LHV models, as
it is possible to use any $\theta\in[0,\pi/2]$, not just the discrete
selection considered above, and the
gap is not generally given by (\ref{gap}). This question is explored further in
\cite{Chistikov2020} and numerically in unpublished work
\cite{Olga2022}.

\begin{figure}
\includegraphics[scale=0.44]{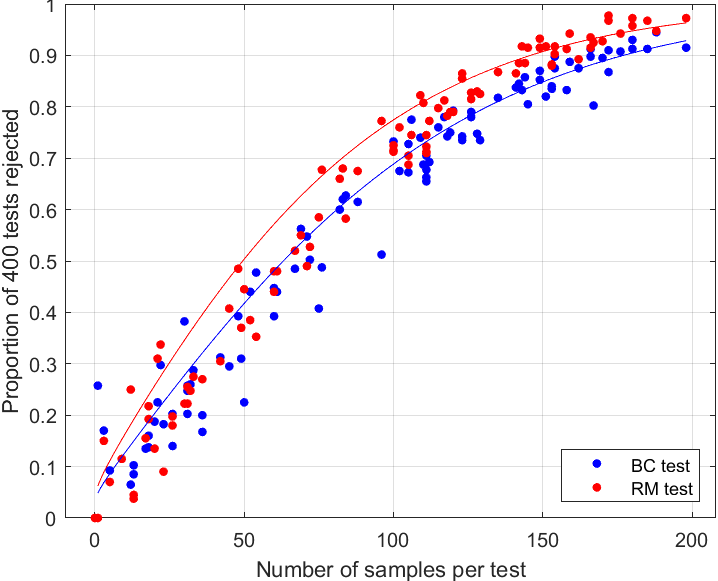}
\centering
\caption{Results when Eve replaces singlets with Bell's LHV model. Test parameters $N=2$ and $\theta=\arcsin(2/\pi)$. Curves are theoretical asymptotic power functions (\ref{asymp}). Dots are empirical data, representing the proportion of 400 simulated tests leading to a rejection of the null hypothesis at the $1\%$ level.}
\end{figure}

However, it is possible to calculate the optimal value of $\theta$ for detecting certain specified classes of LHV models, such as Bell's original model \cite{Bell1964}. A model in this class is defined such that Alice's measurement on one hemisphere of the Bloch sphere gives outcome $+1$ and the other gives $-1$, with Bob's measurement giving opposite values on the same hemispheres. For this model, it is easy to verify that
\begin{equation}
	E[\hat{O}]=-1+\frac{2\theta}{\pi},
\end{equation}
with the difference between this expected correlation and that for the singlet being
\begin{equation}
	\tilde{D}(\theta)=\cos\theta+\frac{2\theta}{\pi}-1.
\end{equation}

This quantity is maximised by $\theta=\arcsin(2/\pi)\approx0.6901$, giving $\tilde{D}(\arcsin(2/\pi))\approx0.2105$.
Hence the random measurement test with this parameter has greatest power for detecting this class of LHV models. For comparison, $\tilde{D}(\pi/4)\approx0.2071$.

Since $\tilde{D}(\theta)>0$ for all $\theta\in(0,\pi/2)$, a test with any $\theta$ in this range would detect Bell's LHV models with some efficiency.

\subsection{Comments on LHV model testing with measurement errors}\label{sec4e}
When Alice and Bob program their measurement devices during a test, there is a possibility they incur small calibration errors. These could be realised as small deviations in their measurement angles on the Bloch sphere. We fix $\delta>0$ as a bound on the magnitude of a deviation in any single measurement for both Alice and Bob.

We examine the effect of such errors on the random measurement and Braunstein-Caves schemes.

\subsubsection{Random measurement scheme}

Theorem 1 of Ref \cite{Kent2013} provides a bound on the expected value of a random measurement sample from an LHV model in this error regime. The theorem equivalently states that for any LHV model, any integer $N\geq2$ and any $\theta\in\left[\pi/2N,\pi/2(N-1)\right)$, we have $-1+1/N\leq E[\hat{O}]\leq1-1/N$.

For $\delta<\pi/8N(N-1)$, it follows that the expected value of a random measurement sample from any LHV model with chosen angle $\theta=\pi/2N+2\delta$ satisfies
\begin{equation}
	|E[\hat{O}]|\leq1-\frac{1}{N}.
\end{equation}

In this setting, the greatest assured difference between the expected correlation for a singlet and that for an LHV model over all possible $\delta$-bounded errors is
\begin{equation}
	D(N)=\cos\left(\frac{\pi}{2N}+4\delta\right)-\left(1-\frac{1}{N}\right).
\end{equation}

$D(N)$ is positive when $\delta<(\arccos(1-1/N)-\pi/2N)/4$, implying that the random measurement test can distinguish between singlet and LHV models in the presence of $\delta$-bounded measurement errors, when $\delta<\min\{\pi/8N(N-1),(\arccos(1-1/N)-\pi/2N)/4\}$.

For $N\geq3$, this required bound on $\delta$ converges monotonically to 0 as $N$ increases. This implies that the scheme can only reliably tolerate a smaller range of absolute measurement errors when $N$ is large, suggesting
that schemes with reasonably large $\theta$ may be more robust.

\subsubsection{Braunstein-Caves scheme}

The expected value of a Braunstein-Caves sample from an LHV model in this error regime is still bounded as
\begin{equation}
	|E[\hat{C}]|\leq1-\frac{1}{N},
\end{equation}
as the Braunstein-Caves inequality holds independently of Alice and Bob's measurement choices.

In this setting, the expected correlation for a singlet over all $\delta$-bounded measurement errors can be calculated using (\ref{bcexp}) and (\ref{bccalc}), by shifting the usual Braunstein-Caves measurement angles for Alice by $\epsilon_A$ and likewise for Bob by $\epsilon_B$, where both $\epsilon_A$ and $\epsilon_B$ represent $\delta$-bounded errors, to give
\begin{equation}
	E[\hat{C}]=-\cos\left(\frac{\pi}{2N}\right)\cos(2(\epsilon_A-\epsilon_B)).
\end{equation}

This implies that the greatest assured difference between the expected correlation for a singlet and that for an LHV model over all possible $\delta$-bounded errors is
\begin{equation}
	D(N)=\cos\left(\frac{\pi}{2N}\right)\cos(4\delta)-\left(1-\frac{1}{N}\right).
\end{equation}

$D(N)$ is positive when $\delta<(\arccos(1-1/N)-\pi/2N)/4$, implying that under this condition, the Braunstein-Caves test can distinguish between singlet and LHV models in the presence of $\delta$-bounded measurement errors.

For $N\geq2$, this required bound on $\delta$ converges monotonically to 0 as $N$ increases. This implies that the scheme can only reliably tolerate a smaller range of absolute measurement errors when $N$ is large,
again suggesting that schemes with small $N$ may be more robust.

\subsubsection{Conclusion}

In summary, it is seen that both schemes are still able to distinguish between singlet and LHV models in the presence of small deviations in the intended measurement angle. As $N$ becomes large, we become less sure of the robustness of each scheme, as the proven range of tolerable measurement errors decreases.

\subsection{Comparison for noisy quantum channel}\label{sec4f}
Consider the scenario in which Eve takes no action, however the quantum channel used for state transmission to Alice and Bob is affected by noise. Different quantum channels are afflicted with different types of noise, however, as a simple example, we can consider a depolarising channel that replaces the singlet with the maximally mixed state with probability $\delta$. 

The effect of this noise on a singlet can be modelled by two-qubit Werner states, with these being the only set of states invariant under arbitrary unitary transformations acting equally on both qubits \cite{Werner1989}. 

The two-qubit Werner state can be defined as
\begin{equation}
	W_\delta=(1-\delta)|\Phi^-\rangle\langle\Phi^-|+\frac{\delta}{4}\mathbb{I},
\end{equation}
where $\delta$ parametrises the strength of the noise, with $\delta=0$ corresponding to a pure singlet state in the absence of noise. 

The expected test samples are calculated in Table \ref{table:3} using (\ref{BCeval}) and (\ref{RMeval}).
\begin{table}[h!]
	\begin{tabular}{ |c|c|c| } 
		\hline
		& $W_\delta$ & $|\Psi^-\rangle\langle\Psi^-|$ \\
		\hline
		$E[\hat{C}]$ & $-(1-\delta)\cos\left(\frac{\pi}{2N}\right)$ & $-\cos\left(\frac{\pi}{2N}\right)$ \\
		\hline
		$E[\hat{O}]$ & $-(1-\delta)\cos\left(\frac{\pi}{2N}\right)$ & $-\cos\left(\frac{\pi}{2N}\right)$ \\ 
		\hline
	\end{tabular}
	\caption{Comparison of expectation values for Werner state and singlet, with the BC test using parameter $N$ and the RM test using $\theta=\pi/2N$.}
	\label{table:3}
\end{table}

Hence, both tests are equally powerful in testing for depolarising noise. Just as in sections \ref{sec4b} and \ref{sec4c}, a larger value of $N$ leads to a test of greater power, so a choice of large $N$ and $\theta=0$ would be optimal.

As an additional example, we can consider the effect of a simple dephasing channel which acts on a qubit as Pauli gate $Z$ with probability $p$. The effect of this noise on a singlet can be described as
\begin{equation}
\begin{split}
\Delta_p&=(1-p)^2|\Phi^-\rangle\langle\Phi^-|\\
&+p(1-p)(\mathbb{I}\otimes Z) |\Phi^-\rangle\langle\Phi^-| (\mathbb{I}\otimes Z)\\
&+p(1-p)(Z\otimes\mathbb{I}) |\Phi^-\rangle\langle\Phi^-| (Z\otimes\mathbb{I})\\ &+p^2(Z\otimes Z) |\Phi^-\rangle\langle\Phi^-| (Z\otimes Z)\\
&=((1-p)^2+p^2)|\Phi^-\rangle\langle\Phi^-|+2p(1-p)|\Phi^+\rangle\langle\Phi^+|,
\end{split}
\end{equation}
where we restrict $0<p<1$.

The expected test samples are calculated in Table \ref{table:4} using (\ref{BCeval}) and (\ref{RMeval}).
\begin{table}[h!]\label{table4}
	\begin{tabular}{ |c|c|c| } 
		\hline
		& $\Delta_p$ & $|\Psi^-\rangle\langle\Psi^-|$ \\
		\hline
		$E[\hat{C}]$ & $-(1-2p(1-p))\cos\left(\frac{\pi}{2N}\right)$ & $-\cos\left(\frac{\pi}{2N}\right)$ \\
		\hline
		$E[\hat{O}]$ & $-(1-\frac{8}{3}p(1-p))\cos\left(\frac{\pi}{2N}\right)$ & $-\cos\left(\frac{\pi}{2N}\right)$ \\ 
		\hline
	\end{tabular}
	\caption{Comparison of expectation values for dephased state and singlet, with the BC test using parameter $N$ and the RM test using $\theta=\pi/2N$.}
	\label{table:4}
\end{table}

It is clear that the random measurement test has greater power in testing for this type of dephasing, for any $0<p<1$. Just as in sections B and C, a larger value of $N$ leads to a test of greater power, so a choice of large $N$ and $\theta=0$ would be optimal.

\section{Discussion}\label{sec5}
While there is no universally superior choice of singlet test, 
we have seen that the random measurement test is theoretically superior or
equal in many natural scenarios, including the detection of
intercept-resend or transformation attacks where Eve prioritises
minimising her chance of detection, distinguishing LHV models, and
detecting rotationally invariant noise.

These results provide a rationale for considering the random
measurement test for singlet verification over more conventional CHSH
schemes (e.g. \cite{McKague2012,Valcarce2019,Bancal2015,Yang2014}).
A complete analysis would consider the full range of attacks open
to Eve and the full range of tests available for A and B.
This would define a two-party game (with A and B collaborating as one party and Eve as the other), in which the
optimal strategy for each party is likely probabilistic.  
However, Eve's actions may be limited depending on how the singlets are generated and
distributed, and on the technologies available to her. 
Also, Alice and Bob may be able to exclude non-quantum LHV attacks
if they can test qubits before measurement
to ensure they are in the appropriate physical state

Our discussion has mainly focussed on the ideal case in which Alice and Bob
can carry out perfectly precise measurements.
Establishing that random measurement tests have an advantage in this case
shows they are potentially valuable options, and motivates 
developing technology that can implement them more easily and precisely.
However, at present, imprecisions need to be taken into account in assessing
the relative feasibility, advantages and costs of all the tests considered.
For example, the Braunstein-Caves test only requires the calibration of measurement
devices in a finite number ($2N$) of orientations around a great circle
on the Bloch sphere, while the random
measurement test requires the ability to measure in all possible
orientations.  The Braunstein-Caves test may thus be a more desirable choice
if calibrating detectors is difficult.

The random
measurement protocol can be implemented in various ways, each of which
costs some resources.   One option is for Alice and Bob to pre-coordinate their measurements.
This requires secure classical communication and/or secure classical memory, albeit not
necessarily a large amount.
For example, if Alice and Bob choose from a pre-agreed list of $10^6$
approximately uniformly distributed axes on the Bloch sphere, they can specify a
measurement pair with about 40 bits, choosing pairs 
separated by the chosen $\theta$ to within error $\lesssim 3 \times
10^{-4}$.
Consuming secure classical communication and/or memory at this rate is
not hugely demanding, and may be a reasonable
option in many quantum cryptographic and communication
scenarios.
However, relatively precise pre-coordinated measurements effectively
define (if pre-agreed) or consume (if securely communicated) large
amounts of shared secret key.   Singlet verification may be required
for only a small fraction of shared singlets.    Still, 
the advantage is at best context-dependent in protocols that aim to generate one-time pads.

An alternative, if Bob has short term quantum memory, is for Alice to communicate
her measurement choice after Bob receives and stores his qubit.
Each can then define their measurement choice by locally generated
or stored random bits, and Bob can delay his measurement choice until he receives Alice's,
with no additional security risk.

Another possible option is for Alice and Bob to choose measurements randomly
and independently, and then sort their results into
approximately $\theta$-separated pairs post-measurement, for 
some discrete set of $\theta$ in the range $[0, \pi/2]$.
This effectively means carrying out random measurement tests
for each $\theta$ in the chosen set, up to some chosen finite
precision.   This protocol effectively uses a random
variable $\theta$, and further analysis is needed to characterise its
efficiency.  The Braunstein-Caves protocol can be similarly adapted to
avoid pre-coordination,
if Alice and Bob each independently choose measurements
from set (\ref{mmset}) and then sorting their results into pairs
that correspond to complete elements of (\ref{mmset}). For a test with parameter $N$, they would on average retain a
fraction $2/N$ of their samples.   If the remainder are discarded,
this requires them to multiply their initial sample size by $N/2$ to
compensate.  However, some of the discarded data could be used for
further Braunstein-Caves tests if N is factorisable.  
Other anti-correlation tests could in principle be used on the
remainder
(althought the finite precision loophole for measurements on
the circle \cite{Chistikov2020} needs to be allowed for).
In the $N=2$ (CHSH) case, there
is no loss of efficiency, as all choices by Alice and Bob would
correspond to an allowed pair.

Larger values of $N$ give more powerful tests for detecting bipartite state transformation
attacks and rotationally invariant noise, while the smallest possible
$N$ is optimal for detecting LHV correlations.  Alice and Bob thus should either
choose $N$ dependent
on which type of attack is most likely or -- if they are in
the type of game-theoretic scenario discussed above -- act against
the potential use of any of the attacks by employing a
probabilistic strategy that mixes different values of $N$.

In the case $N=2$, there is a natural sense in which the random
measurement test is at least as good or better than the
Braunstein-Caves test in every scenario. In section \ref{sec4b}, Eve's goal
could be to carry out an intercept-resend with minimum probability of
detection (i.e. $\psi=0$ or $\beta=0$), in which case the random
measurement test is more powerful. In section \ref{sec4c}, Eve's goal could be
to carry out a state replacement that achieves fidelity $1-\epsilon$
with minimum probability of detection
(i.e. $\langle\Phi^+|\rho_{AB}|\Phi^+\rangle=0$), in which case the
random measurement test is again more powerful. In sections \ref{sec4d} and \ref{sec4f},
both tests are equally good for all variations. In section \ref{sec4e}, it is
seen that both tests are still effective in the presence of small
measurement calibration errors.

Our results thus make a clear case for considering random measurement tests,
and add motivation to ongoing work \cite{Chistikov2020,Olga2022} to identify their power for
the full range of $\theta\in(0,\pi/3)$.
In particular, forthcoming work \cite{Olga2022} provides numerical
evidence on the value of $\theta$ that leads to the greatest expected
difference between correlations resulting from LHV models and
those from a singlet, and the value of said differencce.

Random measurement tests are presently technologically challenging. 
More work is also needed to characterise their robustness in real world applications
where finite precision is inevitable, with various plausible error models,
and where there may be a wide range of plausible adversarial attacks,
against which the optimal testing strategies are likely mixed.
That said, our results suggest that random measurement tests should be considered, as and
when technology allows, in scenarios where efficient singlet testing
is critical and the costs of classical and/or quantum memory resources
are relatively negligible.    

\section*{ACKNOWLEDGMENTS}
The authors acknowledge financial support from the UK Quantum
Communications Hub grant no. EP/T001011/1. G.C. is supported by a studentship from the Engineering and Physical Sciences Research Council. A.K. is partially supported
by Perimeter Institute for Theoretical Physics. Research at Perimeter
Institute is supported by the Government of Canada through Industry
Canada and by the Province of Ontario through the Ministry of Research
and Innovation.

\bibliographystyle{apsrev4-2}
\bibliography{comparingsinglettestingschemes4}

\end{document}